# Etude expérimentale et théorique du mouvement des gouttes immiscibles dans un milieu continu au repos

H. Alla[a], M. Abdelouahab[a], B. Alli Talha[a], F.Z. Beloufa[a]

a. Laboratoire de modélisation et simulation. Département de physique, Université des Sciences et de la Technologie d'Oran, Faculté des sciences, BP 1505 El M'Naour Bir el Djir Oran, 31000, Algérie.

**Résumé :**

*Le problème du mouvement des gouttes dans un milieu continu représente une étape fondamentale pour l'étude des écoulements diphasiques. Leurs domaines d'application sont énormes tels que : la combustion des chambres thermique, extraction d'un liquide d'un autre liquide, catalyseur utilisé pour déclencher des réactions chimiques se produisant en des points d'accès, absorption gazeuse par les gouttes d'eau en météorologie. L'objet de notre travail est de traduire le mouvement des gouttes immiscibles dans l'eau par une relation exprimant la vitesse terminale de la goutte en fonction de son diamètre équivalent. Cette relation bien que théorique a été validée par le model expérimentale et numérique.*

**Abstract :**

*The problem of the movement of the drops in a continuous environment represents a fundamental step for the study of two phases flows. Their domains of application are enormous as: the combustion of the rooms of the machines thermal, extraction of a liquid of another liquid, catalyst used to trigger the chemical reactions occurring in access points, sparkling absorption by the drops of water in meteorology. The object of our work is to translate the movement of the immiscibles drops in water by a relation expressing the terminal velocity of the drop according to her equivalent diameter. This relation although theoretical has been validated by the experimental and numeric model.*

**Mots clefs : vitesse terminale, diamètre équivalent, dynamique des gouttes, milieu continu, simulation numérique.**

## 1 Introduction

Les gouttes sont des objets familiers qui posent nombre de questions, pour le curieux comme pour l'industriel. Parmi les fréquentes situations de deux fluides non miscibles, le cas de la chute d'une goutte dans son environnement (liquide) est séduisant par l'apparition de la configuration de départ : celle d'une goutte initialement sphérique et lâchée dans un fluide au repos, mais le phénomène montre une grande complexité résultant des interactions engendrées par le mouvement relatif des deux milieux qui implique notamment que la goutte se déforme. L'amplitude de cette déformation dépend des grandeurs physiques caractéristiques des fluides en présence, telle que la masse volumique, la viscosité, la tension interfaciale, mais aussi de la grosseur de la goutte. [4]

## 2 Etude expérimentale

Afin d'étudier le mouvement d'une goutte pendante dans un milieu continu, il est nécessaire de déterminer ses paramètres physiques ainsi que la relation entre eux. Un banc d'essais a été réalisé est représenté sur FIG. 1 où l'acquisition des images est faite grâce à une caméra numérique CCD. Les données sont transmises à l'aide d'un ordinateur. Un logiciel de calcul de la géométrie de la goutte traite les différentes images captées.

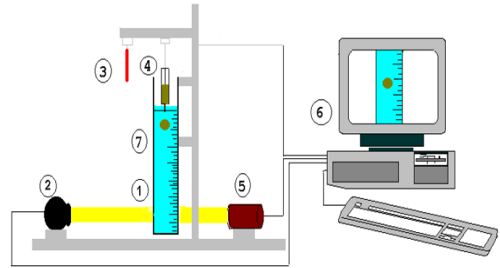

FIG .1- Schéma descriptif du banc de mesure des paramètres physiques d'une goutte fluide pendante dans l'eau. 1-tube en verre, 2-source lumineuse, 3-thermomètre, 4-éprouvette, 5- caméra, 6- appareil d'acquisition, 7-eau.

## 3  Etude théorique et comparaison

Notre étude expérimentale est basée sur une étude théorique [1] décrivant la relation entre la vitesse terminale de la goutte en fonction de son diamètre équivalent.

La courbe théorique de la vitesse terminale en fonction du diamètre équivalent est divisée en deux régions. Cette division se fait par rapport à la goutte de taille minimale ayant une vitesse terminale maximale $U_T$. La première région débutée de zéro jusqu'au pic de la courbe qui représente la vitesse maximale, d'où l'équation (1), [1] cette région correspond aux petite goutte. FIG. 2

$$U_T = \frac{U_L}{\left[1 + \left(\frac{82\,\mu_C\,m_p}{\rho_c\,\alpha_{pm}\,U_L\,d_e}\right)^2\right]^{\frac{1}{2}}} \quad (1)$$

la deuxième commence juste à la fin de la première région et se termine jusqu'à que la goutte atteint son diamètre maximum (diamètre critique) $d_{ecr}$, d'où l'équation (2), [1] cette région correspond aux grosse gouttes.

$$U_T = \left[\frac{g\,\Delta\rho}{4\,\alpha_{pm}\,\rho_c}d_e + \frac{4\,\alpha_{pm}\,\sigma}{\rho_c\,d_e}\right]^{\frac{1}{2}} \quad (2)$$

Où, $U_T$ est la vitesse terminale, $U_L$ est la vitesse limite au pic, $\mu_c$ la viscosité du milieu continu (l'eau du robinet), $\rho_c$ la masse volumique de l'eau, $m_p$ un coefficient arbitraire sans dimension donné au pic, $d_e$ le diamètre équivalent de la sphère, $\sigma$ la tension de surface de la goutte et $\alpha_{pm}$ c'est le rapport de $m_p$ sur m.

En reportant sur une même figure les différentes courbes correspondantes aux équations (1) et (2), sous formes de lignes continues, ainsi que, les mesures faites pour des gouttes d'aniline, de chlorobenzène, de bromobenzène et de carbone disulfure, on obtient ainsi les allures représentées ci-dessous.

Dans un premier temps, notre étude a été faite sur un tube en verre de 10cm de diamètre contenant de l'eau du robinet, FIG. 2, en suite, on a changé de l'eau du robinet en eau distillée sur le même tube. FIG. 4. Dans un tube en verre de 3cm de diamètre, notre étude a été faite pour le seul cas de gouttes d'aniline. On remarque bien les effets de la paroi sur le mouvement des gouttes. FIG. 3

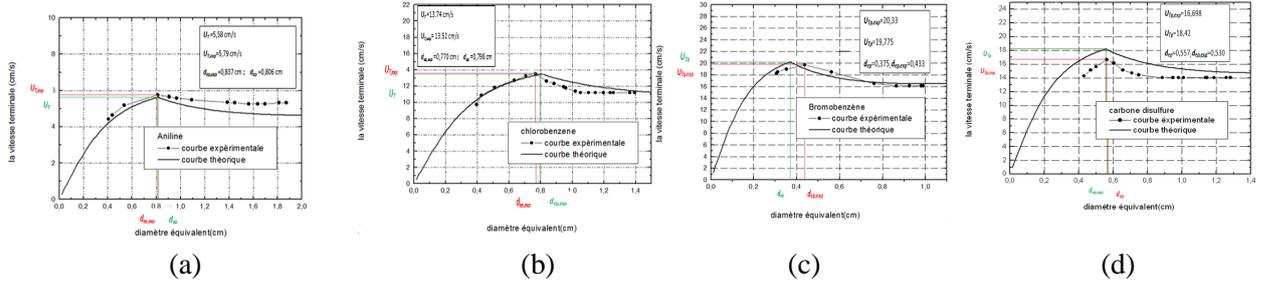

(a)  (b)  (c)  (d)

FIG. 2-comparaison de la vitesse terminale expérimentale avec la vitesse terminale théorique des gouttes en fonction du diamètre équivalent dans un tube de diamètre de 10cm remplis d'eau. (a) courbes d'Aniline, (b) courbe de Chlorobenzène, (c) courbe de Bromobenzène, (d) courbe de Carbone disulfure.

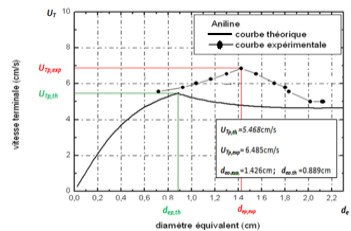

FIG. 3-comparaison de la vitesse terminale $U_T$ expérimentale avec la vitesse terminale théorique des gouttes de l'Aniline en fonction du diamètre équivalent dans un tube de diamètre 3cm remplis de l'eau distillée.

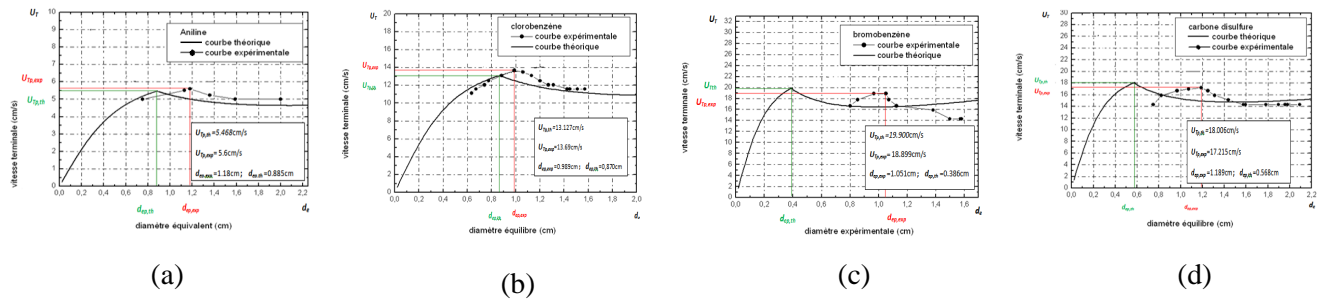

(a)  (b)  (c)  (d)

FIG. 4-comparaison de la vitesse terminale expérimentale avec la vitesse terminale théorique des gouttes en fonction du diamètre équivalent dans un tube de diamètre 10cm remplis de l'eau distillée. (a) courbes d'Aniline, (b) courbe de Chlorobenzène, (c) courbe de Bromobenzène, (d) courbe de Carbone disulfure.

Les courbes expérimentales sont en bon accord avec les courbes théoriques, Les figures représentées ci-dessus montrent la comparaison entre les deux études sur l'évolution de la vitesse terminale de la goutte en fonction de son diamètre équivalent qui ont la même allure ou l'on distingue cinq régions :

Dans La première région nous obtenons la taille minimale de la goutte susceptible de se déformer. La deuxième région correspond à des gouttes de taille moyenne susceptible de se déformer. La troisième région est celle qui correspond à la goutte lorsqu'elle atteint sa vitesse maximale $U_{Tp}$ (limite). C'est la région limite, mais la goutte n'a pas encore atteint la taille maximale. La quatrième région est la région de transition ou la vitesse décroît pendant que le diamètre équivalent croit toujours. La cinquième région est celle où la vitesse devienne quasi-constante, c'est la région critique.

Le tableau suivant montre une étude comparative entre nos valeurs expérimentales et les valeurs théoriques [1] de la vitesse terminale et le diamètre équivalent.

| | Propriétés physico- chimique | | | | | Valeurs théoriques calculées et expérimentales de la vitesse terminale au sommet | | | | | | | | |
|---|---|---|---|---|---|---|---|---|---|---|---|---|---|---|
| Liquide | $\rho_c$ | $\rho_d$ | $\Delta\rho$ | $\mu_c$ | $\sigma$ | $U_L$ | $d_{ep,exp}$ | $\alpha_p$ | $d_{ep}$ | $U_{Texp}$ | $U_{Tth}$ | $K_p$ | $m_p$ | $\alpha_{pm}$ |
| aniline | 0.9981 | 1.016 | 0.0179 | 0.008280 | 6.545 | 6.55 | 0.806 | 0.851 | 0.837 | 5.79 | 5.58 | 1.701 | 3.96 | 4.55 |
| Chlorobenzène | 0.9981 | 1.096 | 0.0979 | 0.008280 | 36.020 | 15.35 | 0.770 | 0.895 | 0.798 | 13.52 | 13.74 | 1.790 | 8.85 | 10.16 |
| Bromobenzène | 0.9971 | 1.488 | 0.4910 | 0.008958 | 37.900 | 23.28 | 0.433 | 0.873 | 0.375 | 19.77 | 20.33 | | 5.82 | 6.69 |
| Carbone disulfure | 0.9991 | 1.260 | 0.2609 | 0.009499 | 45.670 | 20.81 | 0.530 | 0.885 | 0.557 | 16.69 | 18.42 | 1.771 | 7.30 | 8.38 |

TAB. 1- comparaison de valeurs théoriques calculées avec des valeurs expérimentales de la vitesse terminale et du diamètre équivalent au pic.

## 4  Simulation numérique

Dans cette partie, nous avons fait la simulation numérique sous FLUENT 6.2.16 des mouvements des différents gouttes. Fluent est un logiciel industriel qui comporte plusieurs modèles multiphasiques dont le modèle VOF «Volume Of Fluid » qui permet la modélisation de plusieurs phases non miscibles.
Il s'agit donc de simuler l'écoulement dans un cylindre [3] contenant de l'eau, et ouvert à l'air en haut, on injecte des gouttes des fluides (chlorobenzène et l'aniline). Les dimensions et les paramètres du modèle sont les suivantes :-cylindre verticale de hauteur 0.1m et de diamètre 0.02m ; le cylindre est ouvert en haut à la pression atmosphérique , on utilise l'eau et le chlorobenzène aux conditions normale (tension interfaciale 36.020 mN.m$^{-1}$).

Pour des raisons de temps de calcul, nous avons traité le modèle en 2D et pas en axisymétrie. L'idée de cette démarche est d'utiliser deux zones pour les conditions aux limites : l'entrée en haut de type *pressure inlet* ; les parois latérales et en bas de types *wall*. Le modèle sera maillé en éléments *Quad* d'une taille de 0.5mm.

Ainsi, nous obtenons le mouvement de la goutte du chlorobenzène de diamètre 0.4cm à des temps différents. FIG. 5. Le contour des vitesses à des temps différents de la même goutte est illustré en FIG. 6 et le champs des vecteurs de vitesses en FIG. 7.

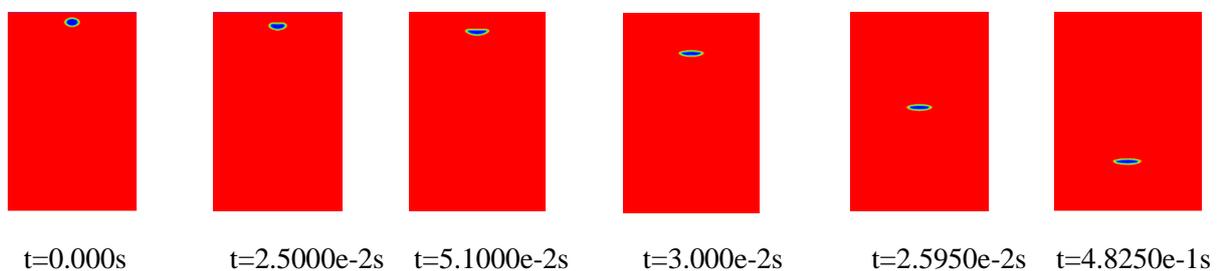

t=0.000s    t=2.5000e-2s    t=5.1000e-2s    t=3.000e-2s    t=2.5950e-2s    t=4.8250e-1s

FIG. 5- contour de volume fraction d'une goutte de Chlorobenzène de diamètre de 0.4cm.

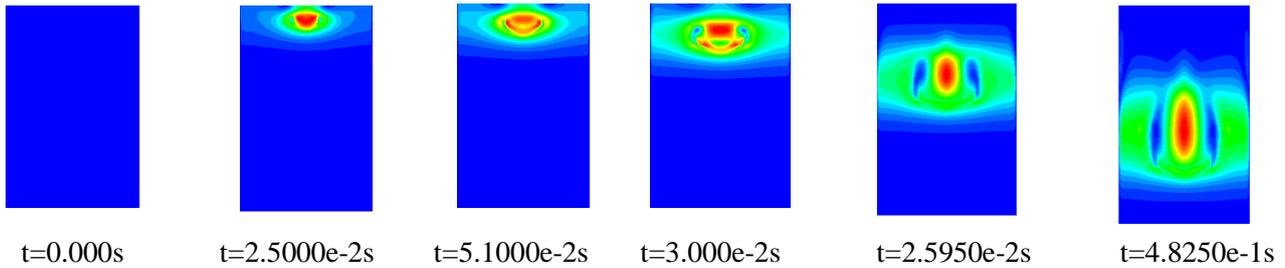

FIG. 6- contour de vitesse moyenne d'une goutte de Chlorobenzène de diamètre de 0.4cm.

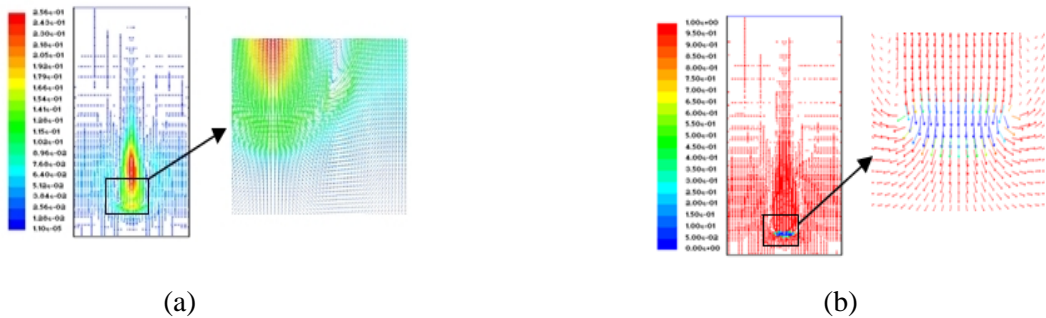

FIG. 7- vecteur de vitesse moyenne(a) vecteur de volume fraction(b) d'une goutte de Chlorobenzène de diamètre de 0.4cm à t=5.53e-1s.

## 5  Résultats et conclusion

Nous avons rassemblé les résultats expérimentaux, théoriques et numériques. FIG. 8, et TAB. 2 pour les deux cas du chlorobenzène et de l'aniline. Nous remarquons très bien que les valeurs des différentes études sont très voisines. Ceci montre bien qu'au lieu de faire des expériences sur de telles produits toxiques qui portent atteintes à la santé, il serait préférable de faire la simulation numérique sous fluent ou autres car les résultats sont en parfaites concordance.

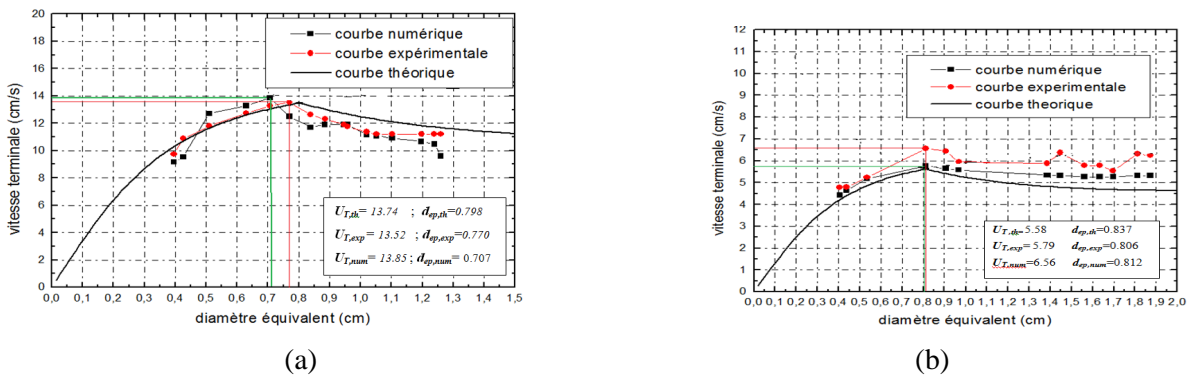

FIG. 8-comparaison des résultats expérimentale et numérique de la vitesse terminale en fonction du diamètre équivalent avec celles théorique. (a) courbe de chlorobenzène, (b) courbe d'aniline.

| Liquides utilisés | valeurs de vitesse terminale et de diamètre équivalent au pic de la courbe | | | | |
|---|---|---|---|---|---|
| | Vitesse terminale théorique $U_{T,th}$ | Vitesse terminale expérimentale $U_{T,exp}$ | Vitesse terminale numérique $U_{T,num}$ | Diamètre équivalent théorique $d_{ep}$ | Diamètre équivalent expérimentale $d_{ep}$ |
| **Chlorobenzène** | 13.74 | 13.52 | 13.85 | 0.798 | 0.770 |
| aniline | *5.58* | *5.79* | 6.56 | 0.837 | 0.806 |

TAB. 2- comparaison de valeurs théoriques calculées avec les valeurs expérimentales et numériques de la vitesse terminale et du diamètre équivalent au pic.

## References

[1] M.Abdelouahab, mouvement des gouttes de liquide dans un milieu continu au repos. Thèse de Doctorat,Oran, Algérie, 2005.
[2] A.Akhtar, V.Pareek, and M.Tadé. CFD Simulations of Continous Flow of Bubbles through Gas-Liquid Columns: Application of VOF Method. Chem. Prod. Proc. Modeling, 2 :9, 2007.
[3]F. Ravelet1, 1 LEMFI - ENSAM,  Introduction `a la simulation sous Fluent d'un problème multiphasique, 2008.
[4] Alla H., Benyettou M., El Ganaoui.M. ,"Numerical investigation of a drop/surface interaction", 13[th] international Heat Transfert ",  Conference ,  13-18 August 2006, sysdney, Australia.